# Modeling of the self-limited growth in catalytic chemical vapor deposition of graphene

**HoKwon Kim**[1], **Eduardo Saiz**[1], **Manish Chhowalla**[2], **and Cecilia Mattevi**[1]*

[1]CASC, Department of Materials, Imperial College London, London,
SW7 2AZ, United Kingdom

[2]Department of Materials Science and Engineering, Rutgers University,
Piscataway,
New Jersey 08854, USA

*E-mail: c.mattevi@imperial.ac.uk

**Abstract**

The development of wafer-scale continuous single-crystal graphene layers is key in view of its prospective applications. To this end, here we pave the way for a graphene growth model in the framework of the Langmuir adsorption theory and two dimensional crystallization. In specific, we model the nucleation and growth of graphene on copper using methane as carbon precursor. The model leads to identification of the range of growth parameters (temperature and gas pressures) that uniquely entails the final surface coverage of graphene. This becomes an invaluable tool to address the fundamental problems of continuity of polycrystalline graphene layers, and crystalline grain dimensions. The model shows agreement with the existing experimental data in the literature. On the basis of the "contour map" for graphene growth developed here and existing evidence of optimized growth of large graphene grains, novel insights for engineering wafer-scale continuous graphene films are provided.

# 1. Introduction

Wafer-scale high quality graphene would enable a variety of applications where exceptionally high electrical conductivity and carrier mobility,[1] outstanding mechanical properties,[2, 3] extremely high impermeability,[4, 5] and chemical inertness are required. In 2009 large area graphene was obtained by CVD on Cu surfaces,[6, 7] and since then significant progress has been made in improving the quality of the CVD grown graphene. Despite the significant effort and advances in the empirical approaches to obtain high quality graphene through CVD, many fundamental aspects of the graphene nucleation and growth process on Cu are not still fully understood. For instance, X. Li et al.[8] and H. Kim et al.[9] have observed that the growth of graphene grains tends to gradually come to a stop especially at low carbon precursor ($CH_4$) concentration or low temperature before complete coverage is reached leaving gaps between the graphene grains even under the continual supply of $CH_4$. Similar observation was also made by C. Kemal et al.[10] for CVD graphene growth on Cu using ethylene ($C_2H_4$) as a carbon precursor instead of $CH_4$. The outcome of the saturated, incomplete coverage is that the grain boundaries exhibit spacing with a dimension spanning from a few nanometers to more than hundreds of nanometers that fragmentize the graphene layer, degrading the electrical transport,[11-13] the mechanical properties,[14] compromising the chemical reactivity.[15, 16] and modifying the permeability.[17] Although the problem of the incomplete final coverage was experimentally overcome by employing a second growth step where the increased concentration of the carbon source in the latter stage of the growth can close the gap yielding a continuous film,[8] the fundamental physical basis for the incomplete saturation graphene area is not yet clear. It deviates from the simple self-limiting mechanism of monolayer graphene growth on Cu surface, since the growth reaction terminates before the entire catalytic surface is fully covered. Kim et al.[9] have postulated, in agreement with prior evidences of similar growth behavior of graphene on Ru,[18] that the saturation area is determined by the difference between the critical supersaturation and equilibrium concentrations of the carbon adspecies. However, a quantitative model to predict the values of supersaturation and equilibrium concentrations and in turn the saturation area for a given set of growth parameters has not been established yet. Therefore, a more detailed, fundamental understanding of the relevant nucleation and growth phenomena in a predictive manner is necessary to achieve a full control over the graphene coverage and quality of the film.

In this article, we present a simple model for the graphene growth by CVD in the presence of methane and hydrogen on Cu surface under the framework of the Langmuir model of competitive adsorption,[19, 20] which has also been previously used to model the decomposition of methane on Ni[20] and homogeneous decomposition of $SiH_4$,[21, 22] and the two dimensional crystallization thermodynamics.[23] Here, we assume surface adsorption up to a monolayer with carbon and hydrogen adatoms mainly comprising the absorbed "2D gas" coexisting with graphene at the surface. The thermodynamic driving forces for dissociative adsorption of $CH_4$ and $H_2$ and formation of graphene were provided by the DFT calculations available in literature in order to predict the rates of the forward and reverse reactions among the vapor ($CH_{4(g)}$ and $H_{2(g)}$), adsorbates ($C_{(a)}$, $H_{(a)}$), and graphene$_{(a)}$ on Cu. The model demonstrates the basic underlying principle that the saturation coverage can be elucidated by considering the reaction equilibrium between vapor and adsorbed phases, and provides a quantitative relationship on the dependence of the final coverage on the growth parameters over an extended range of temperatures and the gas pressures of the reagents.

We would like to underline that the main objective of this study is to start building a foundational work to rationalize the graphene growth with future possibilities for further refinement as more experimental results become available and for adaption to other carbon precursors and metal catalysts, and furthermore extendable to other 2D systems.

**2. Model of Langmuir Adsorption and 2D Crystallization**

Firstly, we give an example of the graphene growth conditions, which lead to maximization the graphene nuclei size but often hinder the possibility to obtain a continuous polycrystalline graphene layer.[8, 9, 11] Figure 1a shows a discontinuous graphene layer with grain lateral sizes larger than ~20 μm on a side on a flat Cu substrate after 40 min of an exposure to $CH_4/H_2$ mixture [The experimental details are given in Supplementary Data (SD)]. Longer growth time did not result in a complete coverage, making the pin-holes between the graphene grains clearly visible under SEM. This saturation of graphene coverage has been observed recently for various conditions at which the effect is more significant for lower growth temperatures. The plot in figure 1b shows the evolution of the graphene area coverage with growth time on the underlying Cu for various growth temperatures with a fixed $CH_4/H_2$ ratio. The saturation values of the graphene area coverage, $A_{sat}$ as a function of growth temperature obtained for a long growth time (> 30 min) are shown in figure 1c. This behavior can be macroscopically explained as following. A slow growth rate, dictated by a low supply of methane in a diluted environment on a very flat Cu surface, tends to minimize the nucleation density[8, 9, 11] but at the same time a balance of the forward and reverse reactions on the Cu surface results in the saturation of graphene coverage before all the graphene grains are connected.[24, 25]

Starting from this observation, we can now try to model graphene growth as a function of temperature and gas partial pressures ($P_{CH4}$ and $P_{H2}$). We use the framework of the modified Langmuir adsorption theory to model the self-limited graphene growth and obtain the saturation coverage. In this context, we first need to consider the balance of chemical surface reactions that lead to the formation of graphene. The overall reaction consists in the conversion of $CH_{4(g)}$ to graphene on Cu surface, $S_{(Cu)}$ and $H_{2(g)}$.

$$CH_{4(g)} + 5S_{(Cu)} \rightleftharpoons Graphene_{(a)} + 2H_{2(g)} \quad (1)$$

This overall reaction can be split into the three individual reversible reactions which lead to graphene formation: 1) the dissociative adsorption of $CH_4$ once it reaches the proximity of the Cu surface and decomposes into monoatomic adsorbates ($C_{(a)}$ and $H_{(a)}$) occupying five Cu surface sites ($S_{(Cu)}$) per one $CH_4$ molecule [equation (2)]; 2) desorption of adsorbed hydrogen, $H_{(a)}$ [equation (3)]; 3) graphene formation from carbon adsorbed ($C_{(a)}$) onto the Cu surface [equation (4)].

$$CH_{4(g)} + 5S_{(Cu)} \underset{k_{-1}}{\overset{k_{+1}}{\rightleftharpoons}} C_{(a)} + 4H_{(a)} \quad (2)$$

$$4H_{(a)} \underset{k_{-2}}{\overset{k_{+2}}{\rightleftharpoons}} 2H_{2(g)} + 4S_{(Cu)} \quad (3)$$

$$C_{(a)} \underset{k_{-3}}{\overset{k_{+3}}{\rightleftarrows}} \text{Graphene}_{(a)} \quad (4)$$

On the basis of the observation that the direct interaction of CH$_{4(g)}$ and H$_{2(g)}$ with graphene surface are not likely to occur, as both the decomposition of CH$_{4(g)}$ onto graphene and the etching of graphene by H$_{2(g)}$ without the metal catalyst are found to be negligible in the typical range of deposition temperatures (700 – 1050 °C)[11, 26]; we consider only the adsorption and desorption of these species at the Cu surface. Often the CVD of graphene is performed in the presence of excess H$_2$ as it provides reducing atmosphere to prevent the oxidation of Cu, however, etching of graphene on Cu is expected by the adsorbed hydrogen [reverse reaction of equation (2)] if the flow of CH$_4$ into the system is stopped.[27] Furthermore, we rule out the effect of Cu evaporation that was reported to be detrimental for complete coverage in some LPCVD experiments[28, 29] This does not appear to be significant in our case as less saturation area coverage was observed for lower growth temperature where the evaporation rate of Cu should be negligible.

We note that the CH$_4$ decomposition reaction can be broken down further into more steps involving the intermediate species that may form on Cu. Several contributions[20, 30] consider the intermediate steps of equation (2) on transition metal surfaces in the usual growth conditions to be the following:

$$CH_{4(g)} + S_{(Cu)} \underset{k_{-1a}}{\overset{k_{+1a}}{\rightleftarrows}} CH_{4(a)} \quad (5)$$

$$CH_{4(a)} + S_{(Cu)} \underset{k_{-1b}}{\overset{k_{+1b}}{\rightleftarrows}} CH_{3(a)} + H_{(a)} \quad (6)$$

$$CH_{3(a)} + S_{(Cu)} \underset{k_{-1c}}{\overset{k_{+1c}}{\rightleftarrows}} CH_{2(a)} + H_{(a)} \quad (7)$$

$$CH_{2(a)} + S_{(Cu)} \underset{k_{-1d}}{\overset{k_{+1d}}{\rightleftarrows}} CH_{(a)} + H_{(a)} \quad (8)$$

$$CH_{(a)} + S_{(Cu)} \underset{k_{-1e}}{\overset{k_{+1e}}{\rightleftarrows}} C_{(a)} + H_{(a)} \quad (9)$$

However, the kinetic parameters of the intermediate steps can be combined to give effective an equilibrium constant for the overall reaction (details are given in SD), considering that the three reactions are sufficiently independent.

As of now, there is no conclusive answer whether the main active species for the formation of graphene on Cu are carbon adatoms,[31, 32] hydrogenated carbon species,[33, 34] or carbon clusters.[32, 35] Here, for simplicity and for the fundamental demonstration of our model, we consider that the carbon adsorbed onto the Cu surface is in the form of adatoms, and given the flexibility of our model, it can readily be refined in the light of future experimental evidence.

On the basis of the existing experimental data and DFT calculations available in the literature, we can estimate that the enthalpy of adsorption of methane on Cu is about 3.2 eV,[27, 36, 37] the hydrogen adsorption enthalpy is about -0.3 eV,[38-40] and the enthalpy of graphene formation from adsorbed carbon is about -2.4 eV.[32, 34]

We then express the rate of adsorption of methane or hydrogen according to the gas kinetic theory as:

$$r_{ad} = \frac{P}{\sqrt{2\pi mkT}} s_0 \exp\left(\frac{-E_{ad}}{kT}\right) f(\theta_s) \quad (10)$$

where $P$ is the partial pressure of a reactant, $m$ is the mass of a gas molecule, $k$, Boltzmann's constant, $T$, temperature, $s_0$ is the initial sticking coefficient pre-exponential factor, and $f(\theta_s)$ is free surface coverage ($\theta_s$) dependent sticking coefficient term, $E_{ad}$ activation energy of the adsorption.

Similarly, the rate of desorption of any adsorbed species on substrate sites can be expressed as:

$$r_{des} = v_n \exp\left(\frac{-E_{des}}{kT}\right) [A]^n \quad (11)$$

where n, is the order of the reaction, $v_n$ is n-th order vibrational frequency and we assume $v_n = 10^{13}$ s$^{-1}$ for most of the cases, which is a generally used value of a vibrational frequency when the experimental value is not known, [A] concentration of the adsorbed species, $E_{des}$, activation energy of desorption.

Therefore, the rates of adsorption ($r_{+1}$) and desorption ($r_{-1}$) for the methane decomposition reaction [equation (2)] are respectively:

$$r_{+1} = k_{+1} P_{CH_4} [S_{Cu}]^5 \quad (12),$$

and $r_{-1} = k_{-1}[C][H]^4 \quad (13)$

Now, balancing rates of adsorption and desorption for the CH$_4$ decomposition reaction at equilibrium, we can define the constant of equilibrium, $K_1 = k_1/k_2$ for equation (2) as:

$$K_1 = \frac{1}{\rho_s v_1 \sqrt{2\pi m_{CH_4} kT}} \exp\left(\frac{-\Delta H_{ad\_CH_4}}{kT}\right) = \frac{(\theta_C)(\theta_H)^4}{P_{CH_4}(\theta_S)^5 (1-\theta_G)^5} \quad (14)$$

Typically surface concentrations without the presence of graphene can be expressed as [C] = $\theta_C \rho_s$ and [H] = $\theta_H \rho_s$ where $\rho_s$ is the density of the surface sites on Cu (~1.5 × 10$^{19}$ m$^{-2}$). However, here we assume that C and H adatoms competitively bind only to the available surface sites that are not covered by graphene such that [C] = $\theta_C \rho_s$ /(1 - $\theta_G$) and [H] = $\theta_H \rho_s$ /(1 - $\theta_G$). In addition, the sticking coefficient according to the Langmuir theory of monolayer adsorption is assumed to be[19]:

$$s_0 f(\theta_s) = \theta_s = 1 - \theta_H - \theta_C - \theta_G \quad (15)$$

We note that the coverage by CH$_x$ species resulting from the decomposition of methane are not considered since they are considered to be short-lived.[30]

Similarly for H$_2$ desorption reaction [equation (3)], the rates of desorption and adsorption are:

$$r_{+2} = k_{+2}[H]^4 \quad (16)$$

$$r_{-2} = k_{-2}(P_{H_2})^2 (\theta_S \rho_s)^4 \quad (17)$$

Equating the two rates to obtain the equilibrium constant,

$$K_2 = \rho_s^2 v_2^2 (2\pi m_{H_2} kT) \exp\left(\frac{2\Delta H_{ad\_H_2}}{kT}\right) = \frac{(P_{H_2})^2 (\theta_S)^4 (1-\theta_G)^4}{(\theta_H)^4} \quad (18)$$

Finally, for graphene formation reaction, equation (4), we consider the balance of attachment and detachment rates of carbon atoms per unit length of graphene phase boundary based on two-dimensional crystallization kinetics.[19]

$$r_{+3} = k_{+3}[C] = a_{cu} v_{+3\_Cu} \exp\left(\frac{-E_{att}}{kT}\right)[C] \quad (19),$$

and $r_{-3} = k_{-3} = \dfrac{v_{-3\_G}}{a_G} \exp\left(\dfrac{-E_{det}}{kT}\right)$ (20)

where $v_{+3\_Cu}$ and $v_{+3\_G}$ are the vibrational frequency factors for Cu and graphene, respectively, and $a_{Cu} = 2.3 \times 10^{-10}$ m and $a_G = 1.42 \times 10^{-10}$ m are the lattice spacing for Cu and graphene, respectively.

The equilibrium constant for the reaction is then:

$$K_3 = \dfrac{a_{cu} a_G v_{+3\_Cu}}{v_{-3\_G}} \exp\left(\dfrac{-\Delta H_{form\_G}}{kT}\right) = \dfrac{1-\theta_G}{\theta_C \rho_s} \quad (21)$$

Solving for the coverage of graphene from above equations, we obtain:

$$\theta_G = 1 - \dfrac{K_2^{\frac{1}{4}} \left(P_{H_2}\right)^2}{K_1 K_2^{\frac{5}{4}} K_3 P_{CH_4} \rho_s - K_1 K_2^{\frac{5}{4}} P_{CH_4} - \left(P_{H_2}\right)^{\frac{5}{2}}} \quad (22)$$

In the typical experimental conditions ($P_{CH4}$, $P_{H2}$ < 1 MPa, T = 300 K – 1080 K), $\theta_S \gg \theta_C, \theta_H$. Thus,

$\dfrac{\left(P_{H_2}\right)^2}{K_1 K_2 K_3 \rho_s P_{CH_4}} = \theta_S$ is the dominant term giving rise to exponential behavior with apparent activation energy of $\Delta H_{ad\_CH4} - 2\Delta H_{ad\_H2} + \Delta H_{form\_G}$.

Rearranging equation (22), we obtain the overall equilibrium constant, $K_G$, for the conversion of methane (CH4$_{(g)}$) to graphene adsorbed on Cu [equations (2)+(3)+(4)]:

$$K_G = K_1 K_2 K_3 = \dfrac{\left(P_{H_2}\right)^2}{P_{CH_4} \theta_s \rho_s} = \dfrac{\left(P_{H_2}\right)^2}{P_{CH_4} [S_{Cu}]} \quad (23)$$

where the graphene coverage can also be written as:

$$\theta_G \approx 1 - \dfrac{\left(P_{H_2}\right)^2}{K_G \rho_s P_{CH_4}} \quad (24)$$

We now have an expression for the area coverage of graphene ($A_{sat} = \theta_G$) as a function of the temperature [equation (22)] and we can fit this to the experimentally obtained graphene coverage values. Fixing the enthalpy values and using the pre-exponential factor of $K_3$ as the only fitting parameter, we obtained the value of the pre-exponential coefficient from the curve fitting to be 6.3 x $10^{-18}$ $m^{-2}$ (figure 1c). This is a reasonable value as the vibrational frequency factors can vary over a few orders of magnitude.[23] Using one of the energy values as an additional variable results in the change of only ~14% of the original value which is within the range of predicted values for Cu(100) and Cu(111).

Based on the calculated values of equilibrium constants, we can identify the set of growth conditions for a continuous coverage of graphene, incomplete coverage of graphene, and no-possible formation of graphene. In figure 2, the contour plot for graphene growth is shown for extended range of experimental conditions of temperature and the ratio, $P_{CH4}/P_{H2}^2$, where we can assume that the enthalpy values and pre-exponential factors remain constant. Moreover, various experimental data points have been added in order to test our model in the light of the experimental results already published in the literature. In the region with zero graphene coverage, $\theta_G < 0$ (figure 2, white area), the adsorbed carbon remains as two dimensional gas without forming a $sp^2$ network as the surface concentration of carbon adatoms is low.[19] Between $\theta_G = 0.997$ and $\theta_G = 0$, the ensemble of growth parameters leads to the coexistence of graphene islands and adsorbed carbon in the copper surface without actually enabling the formation of continuous coverage. Note that $P_{CH4}/P_{H2}^2$ was chosen as the main parameter for the gas phase because $\theta_G$ is virtually independent of individual gas partial pressures if the ratio remained fixed [equation (24)].

We can see that both sets of the data points related to continuous graphene layer[6, 11, 16, 29, 41-56] and discontinuous graphene layer[45, 46, 50, 53, 55, 57-59] generally fall within the reasonable range of $P_{CH4}/P_{H2}^2$ values and temperature predicted to provide the same behavior. Further, we have also plotted our experimental data (red squares) referring[9] to graphene grown at different temperatures where the graphene coverage can be either complete or incomplete depending on the temperature. The possible sources of discrepancy between our predictions and the experimental data are discussed in the discussion section.

## 3. Degree of Supersaturation and Chemical Potential

Our theory can be further applied to provide a useful insight regarding the degree of supersaturation during graphene nucleation as the saturation area coverage can also be determined by the supersaturation concentration of adsorbed carbon, [C]$_{sup}$ at the onset of nucleation.[19] On the basis of Langmuir theory, the supersaturation concentration of carbon adatoms can be estimated to be [C]$_{sup}$ = $\theta_c' \rho_s$ where $\theta_c'$ is the coverage of carbon atoms considering only the balance of equations (2) and (3) on Cu surface before the onset of graphene nucleation and growth. This sets upper limit on the physical supersaturation level of carbon adatoms. Within the limit of attachment/capture controlled nucleation and growth, this gives a reasonable upper limit for the [C]$_{sup}$ since adsorption and desorption equilibrium [equations (2) and (3)] is reached more quickly than the attachment/detachment equilibrium [equation (4)]). For a finite graphene coverage, [C]$_{sup}$ must be greater than the equilibrium carbon concentration, [C]$_{eq}$ = $1/K_3$. Figure 3a shows the variation of [C]$_{sup}$ and [C]$_{eq}$ as a function of temperature. The intersection (indicated by an arrow) determines the minimum growth temperature required to form graphene. Similar temperature dependent behavior for [C]$_{sup}$ and [C]$_{eq}$ has been also reported for graphene growth on Ru in ultra-high vacuum conditions.[18, 60, 61]

Following from equation (24), the equation for saturation area coverage can be also written as:

$$\theta_G \approx 1 - \frac{[C]_{eq}}{[C]_{sup}} \qquad (25)$$

This is because,

$$\frac{[C]_{eq}}{[C]_{sup}} = \frac{1/K_3}{\theta_C' \rho_s} \qquad (26)$$

$$\theta_C' = \frac{K_1 K_2 P_{CH4} \theta_s'}{\left(P_{H_2}\right)^2} \qquad (27)$$

$\theta_s' = 1 - \theta_C' - \theta_H' \approx 1$ as $\theta_s' \gg \theta_C', \theta_H'$ again in the typical range of conditions.

The supersaturation chemical potential, $\mu_{sup} = kT\ln([C]_{sup}/[C]_{eq})$ represents the driving force to form graphene at the onset of nucleation and it can be conveniently used as criteria for nucleation and growth of graphene for a set of various growth conditions. We have calculated the supersaturation chemical potential values for the data points based on the reported values of partial pressures of $CH_4$, $H_2$ and growth temperatures as illustrated in figure 3b.

## 4. Discussion

The deviations between our predictions in the computed contour map (figure 2) and the observed range of the experimental data points may arise from various sources. First, the inaccuracies in the condition-independent enthalpy values extrapolated by the DFT calculations and the equilibrium pre-exponential factors due to variability of the surface morphology and crystallinity of the Cu surface may introduce an uncertainty in predicting the coverage in other experimental works, In addition, due to large differences between the flow conditions and designs of the experimental systems employed by the published works, it may be difficult to identify the accurate partial pressure values across the substrate surfaces. For example, the most commonly used Pirani or ionization gauges for pressure measurement are not gas independent and not reliable for measuring the pressure of gas mixtures without precise calibration for low pressure (LPCVD) experiments. In our experiment, we have specifically used a gas-independent capacitance gauge near the sample stage in order to monitor the actual gas pressures as accurately as possible. Moreover, especially for APCVD growth experiments, where the mean free path of a gas particle is short, the growth can be diffusion limited, and a thick boundary layer can develop across the substrate.[43, 62, 63] This can impart gas compositions at the surface which is different from that of the bulk-flow. Lastly, for high temperature and pressure conditions, the homogenous, non-catalytic decomposition of $CH_4$ can also occur.[64-66] This may be able to explain the larger-than-expected coverage of graphene for the significant deviations evident in 4 APCVD data points[52, 56-58] out of 37 in total. We have associated uncertainties to some of the possible sources of error (in enthalpy, pre-exponential exponents, and pressure estimates) and computed the error bounds for the contour map in figure S1 in SD. It is worth noting that the errors are not significant (below one order of magnitude) and that about 4/5 of the data points from the literature fit well with our model within the range of the error bounds.

Another possible challenge to our model is that in the limit of high $\theta_G$ close to 1, there may be an additional energy required to properly "stitch" the graphene grains of different orientations to for a continuous polycrystalline graphene. This may be especially true for Cu(100) and other high index planes where rotated domains of various misorientation angles are frequently observed. Since the formation energies for the grain boundaries are so far unknown, we assume that the graphene grains exhibit a single rotational orientation as in the case of Cu(111) or that the grain boundary energies are negligible. In the limit of low supersaturation, $\mu_{sup} \sim 0$, we have not considered the additional energy barrier to form the critical nuclei whose size has not been reliably predicted so far. $\theta_G$ value estimated by our model in this limit represents the upper limit where there is no barrier for nucleation. Further systematic experiments will help to address these issues and refine the predictions.

Overall, on the basis of our model, it is generally advisable to perform graphene growth at high partial pressure of the carbon source and high growth temperature in order to obtain a continuous layer. High temperature is also beneficial to decrease the density of nuclei, which is also in agreement with experimental observations[8, 9] and the theoretical predictions of the rate equations model.[9, 67] However, graphene grains of significantly larger size are formed under the extremely low pressure conditions in which continuous graphene coverage cannot be achieved.[8, 11] To overcome this hurdle, one possible solution is employing two-step growth where large grains of graphene nucleate at low density and grow to saturation under low $P_{CH4}$ and high $P_{H2}$ in the first step, and then $P_{CH4}$ is increased in order reach a continuous graphene in the 2$^{nd}$ stage. Indeed this has been experimentally shown by X. Li et al.[8, 25] after empirical observation that the density of nuclei increases at lower temperatures. Now to do so, appropriate gas pressures can be chosen by directly referring to our contour plot.

Another practical implication of our model for improving the quality of the CVD graphene is that fast cooling and termination of $H_2$ flow into the system at the end of the growth[29] are advised for defect free graphene coverage. This is because the chemical potential toward complete coverage of graphene decreases for fixed $P_{CH4}/P_{H2}$ which then leads to the catalytic etching of graphene by $H_2$ during cooling. This effect of $H_2$ etching has also been observed by Zhang et al.[27] who have shown that etching becomes more significant at lower temperature than 1000 °C as the equilibrium shifts toward the left side of the $CH_4$ decomposition reaction [equation (2)]. Therefore, the conditions for continuous graphene must be preserved during cooling by adjusting $P_{CH4}/P_{H2}$ or the system must be rapidly cooled to prevent the etching of graphene.

Moreover, in the framework of our model, we can also estimate the concentration of C adatoms based on the experimental results of area coverage evolution over time. Previously, the expression for time-dependent graphene area coverage, $A_G(t)$ can be obtained by solving the following differential equation of a simple, edge-controlled kinetics:[9]

$$\frac{dA_G}{dt} = k_{att} c_{cu} \sqrt{A_G} - k_{det} \sqrt{A_G} \qquad (28)$$

where $k_{att} c_{cu} \sqrt{A_G}$ are the rate of graphene area coverage increase due to atoms arriving, that are proportional to the concentration of adsorbed atoms on the graphene-free Cu surface and to the perimeter of the graphene island ($\sqrt{A_G}$) and $k_{det} \sqrt{A_G}$ are the rate of decrease in the area coverage due to atoms leaving.

The equation (28) can be modified according to our model that:

$$c_{cu}(t=0) = c_{nuc} = [C]_{sup} \qquad (29)$$

assuming that adsorption and desorption equilibrium is reached much faster than the nucleation of and growth of graphene, $c_{nuc} = [C]_{sup}$ at the onset of the nucleation.

And $c_{cu}(t \to \infty) = c_{eq} = [C]_{eq}$ (30)

Using the relationship for the predicted saturation area, $\theta_G$, $[C]_{sup}$ and $[C]_{eq}$:

$$A_G(t \to \infty) = A_{sat} = \theta_G \approx 1 - \frac{[C]_{eq}}{[C]_{sup}} \quad (31)$$

Also, $\dfrac{dA_G(t \to \infty)}{dt} = 0 \Rightarrow k_{att} c_{nuc}(1-\theta_G) = k_{det}$ (32)

Substituting the above relationships to equation (28) and solving the equation, we can write the evolution of the graphene area coverage as:

$$A_G = \theta_G \left( \frac{e^{F(t-t_0)}+1}{e^{F(t-t_0)}-1} \right)^2 \quad (33)$$

where $F = k_{att}\sqrt{\theta_G} = \dfrac{k_{+3} c_{nuc} \sqrt{N_s \theta_G}}{\rho_G} = \dfrac{a_{cu} v_{+3\_Cu}}{\rho_G} \exp\left(\dfrac{-E_{att}}{kT}\right) c_{nuc}\sqrt{N_s \theta_G}$.

Here, $k_{att}$, which gives the rate of the overall fractional area increase per unit time, can be linked to $k_{+3}$ which gives the number atoms arriving per unit length of a nucleus edge per unit time, as $\dfrac{da_G}{dt} = \dfrac{k_{+3}[C]}{\rho_S}\sqrt{a_G}$ and $A_G = N_s a_G$, where $a_G$ is a mean area of graphene nucleus, and $N_s$ is the density of nuclei independent of time assuming instantaneous nucleation.

Therefore, $k_{att} = \dfrac{k_{+3}\sqrt{N_s}}{\rho_G}$ (34)

Then the exponential factor, F, related to the rate constant of carbon attachment now becomes:

$$F = \frac{k_{+3} c_{nuc}\sqrt{N_s \theta_G}}{\rho_G} = \frac{a_{cu} v_{+3\_Cu}}{\rho_G}\exp\left(\frac{-E_{att}}{kT}\right) c_{nuc}\sqrt{N_s \theta_G} \quad (35)$$

Then, using the experimentally obtained F, and $N_s$, and $\theta_G$, and estimated value of $k_{+3}$, $c_{nuc}$ can be calculated by the following equation:

$$c_{nuc} = \frac{\rho_G F}{k_{+3}\sqrt{N_s \theta_G}} \quad (36)$$

The calculation for $c_{nuc}$ based on the analysis of the experimental values of $F$, $N_s$, and $\theta_G$ (figure 4) yields the supersaturation surface carbon concentration values on the order of $\sim 1 \times 10^9$ m$^{-2}$ in the range of 720 – 1000 °C, which are remarkably similar to the range of $[C]_{sup}$ values predicted by our model, although the large uncertainty in $v_{+3\_Cu}$ and F makes further analysis difficult. This $[C]_{sup}$ value is much lower than surface carbon adatom concentration of $> 10^{16}$ m$^{-2}$ on Ru(0001)[68] (consequence of high adsorption energy of C on Ru) that has been measured by in-*situ* LEEM. The extremely low surface concentration is possibly the reason that the recent attempt to directly measure the surface carbon concentration on Cu by LEEM or XPEEM techniques[69] has been fruitless so far.

We would like to underline that our basic approach based on the balance of adsorption/desorption and surface reactions can be extended to consider different active species (e.g. $CH_x$, carbon clusters, and carbon chains) and some of these calculations are underway. However, this will not detract from the main conclusion (the existence of three regions defined by the balance of the reactions). Furthermore, the fact that we can obtain a good predictive fitting by considering only C adatom species may suggest that these species play an important contribution.

## 5. Conclusions

Summarizing, we have modeled the graphene growth by CVD on Cu surfaces to formulate the criteria for final graphene coverage. The latter provides the range of growth parameters (temperature and gas pressures) in which two phases (graphene and adsorbed carbon) define the three fields: a continuous polycrystalline graphene layer, an uncompleted polycrystalline graphene layer coexisting with adsorbed carbon, and adsorbed carbon species. The "phase map", corroborated by experimental data, provides strategies to address the fundamental problems of continuity, and crystalline grain dimensions at the same time. The model is versatile and it can be extended to different carbon precursors, from solid to liquid phase as well as different catalysts and different temperature ranges. Extension to the growth of other 2D atomically thin materials, such as the newly emerging transition metal dichalcogenides (TMDs), and boron nitride will also be of particular interest.

**Acknowledgements**

This project was financially supported by the NSERC (Canada), The Leverhume Trust, and Imperial College London. C.M. acknowledges the award of The Royal Society Univeristy Research Fellowship by the U.K. Royal Society Research.

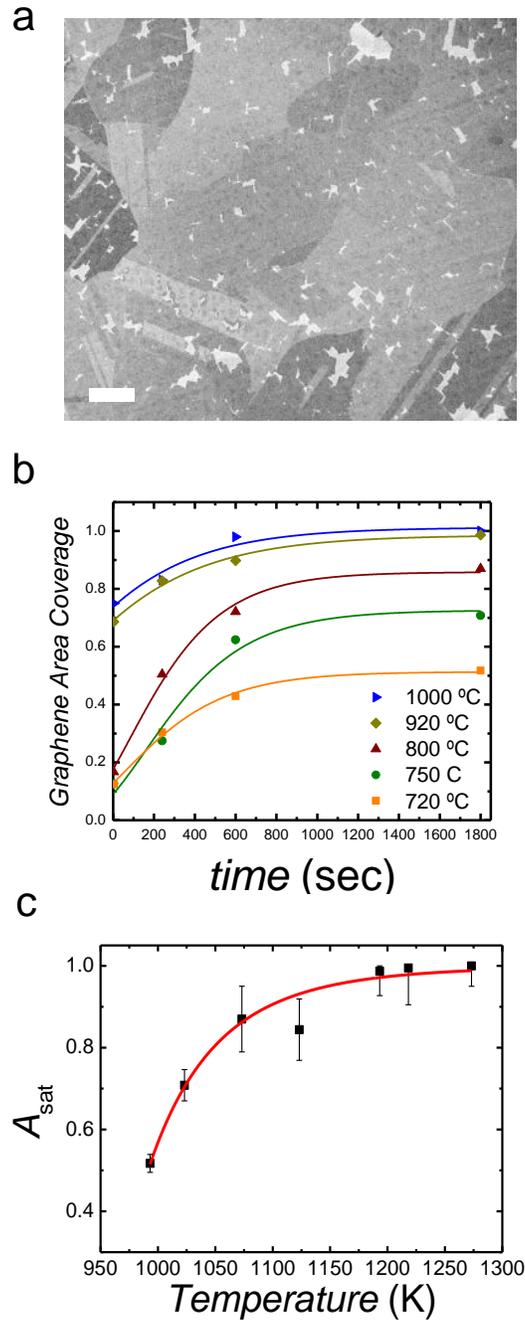

**Figure 1.** (a) SEM image of graphene grown on Cu at saturation. At reduced methane partial pressure, a full complete coverage of graphene cannot be achieved leading to pin-holes and cracks in the layer (visible as bright regions of exposed Cu). Scale bar: 100 μm. (b) Temperature dependent graphene area coverage versus time with the curve fitting using the edge-controlled kinetics of graphene formation.[9] Below 1000 °C, the final graphene coverage is self-limited to a saturation value, $A_{sat} < 1$. (c) Saturation graphene coverage ($A_{sat}$) versus growth temperature. The curve fitting was performed by equation (22).

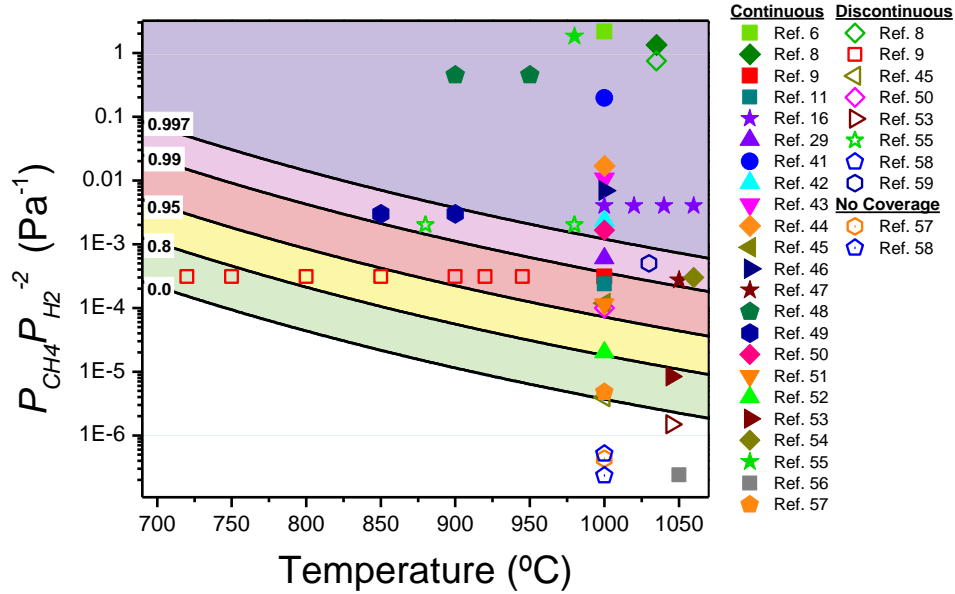

**Figure 2.** Contour plot for saturation graphene coverage, $\theta_G$ ($\theta_G$ values are indicated by the labels near the left axis). Values calculated from experimental conditions reported in the literature for various growth conditions have been plotted for comparison.

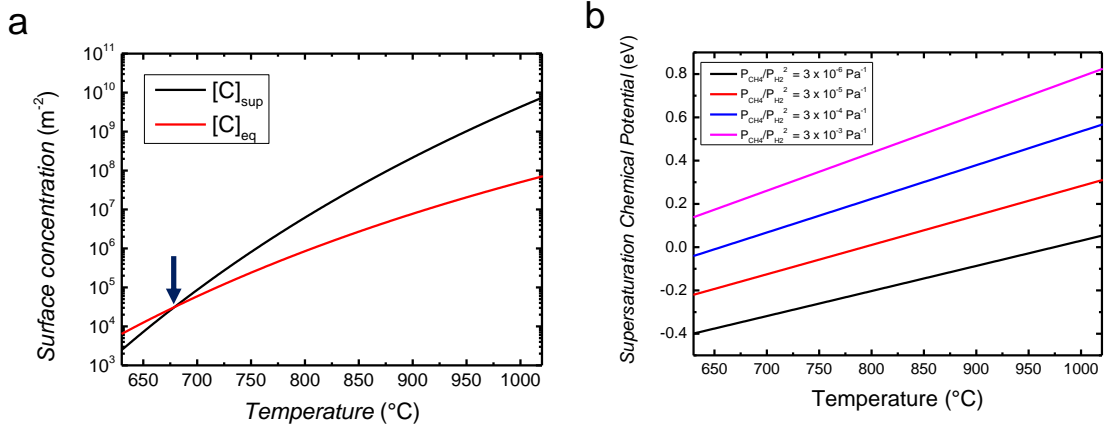

**Figure 3.** (a) Variation of supersaturation carbon concentration and equilibrium carbon concentration with temperature. The arrow indicates the temperature (~954 K) below which graphene formation does not occur as $[C]_{sup} < [C]_{eq}$. The $CH_4$ and $H_2$ partial pressures used for the calculation are $P_{CH4}$ = 40 Pa and $P_{H2}$ = 360 Pa. (b) Variation of supersaturation chemical potential with partial pressure ratios and temperature.

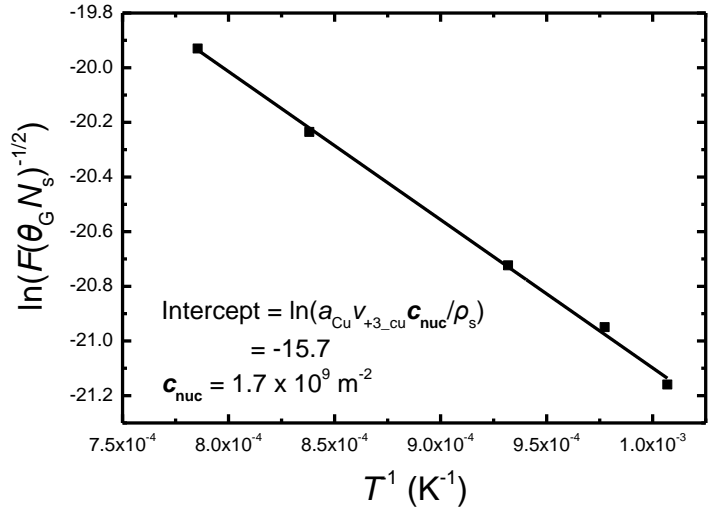

**Figure 4.** $\ln(F(\theta_G N_s)^{-1/2})$ vs. 1/T curve based on the temperature dependent experimental growth kinetics. The temperature dependent values of F were calculated from the fitted values of time dependent graphene area coverage in figure 1b based on equation (33). The linear fitting was performed to estimate the value of critical supersaturation concentration, $c_{nuc}$ using equation (35).

# Supplementary Data

**Title: Modeling of the self-limited growth in catalytic chemical vapor deposition of graphene**

**HoKwon Kim[1], Eduardo Saiz[1], Manish Chhowalla[2], and Cecilia Mattevi[1]***

[1]CASC, Department of Materials, Imperial College London, London, SW7 2AZ, United Kingdom

[2]Department of Materials Science and Engineering, Rutgers University, Piscataway, New Jersey 08854, USA

*E-mail: c.mattevi@imperial.ac.uk

### Experimental Details

*Chemical vapor deposition of graphene on Cu*: The experimental setup for the chemical vapor deposition growth of graphene is described in previous reports.[1] In short, we have employed 0.025 mm thick copper foil of 99.8 at. % purity (Alfa Aesar #13382 lot H18W024) for the Cu substrate. For the high methane partial pressure growth ($P_{CH4}$) to reach a complete coverage at 1000 °C (Figure 1c and 1d), 0.4 mbar of $P_{CH4}$ and 3.6 mbar of hydrogen partial pressure ($P_{H2}$) were used. For the low $P_{CH4}$ growth (Figure 1a), where the complete graphene coverage was not achieved, $P_{CH4}$ and $P_{H2}$ were 0.0015 mbar and 0.2 mbar, respectively.

*Graphene saturation area measurement*: The experimental saturation area ($\theta_G$) was measured by analyzing the SEM images via ImageJ software greyscale analysis for graphene samples grown at different temperatures for long growth times (> 30 min).

## Model details

Consider three main reversible reactions that result in conversion of $CH_{4(g)}$ to graphene formed on Cu and $H_{2(g)}$:

$$CH_{4(g)} + 5S_{(Cu)} \rightleftharpoons Graphene_{(a)} + 2H_{2(g)}$$

(1) Dissociative adsorption of methane:
$$CH_{4(g)} + 5S_{(Cu)} \underset{k_{-1}}{\overset{k_{+1}}{\rightleftharpoons}} C_{(a)} + 4H_{(a)} \quad \Delta H_{ad\_CH4} \cong 3.2 \text{ eV}^{[2]}$$

(2) Hydrogen Desorption:
$$4H_{(a)} \underset{k_{-2}}{\overset{k_{+2}}{\rightleftharpoons}} 2H_{2(g)} + 4S_{(Cu)} \quad \Delta H_{ad\_H2} \cong -0.3 \text{ eV}^{[3]} \text{ per } H_2$$

(3) Graphene Formation from adsorbed carbon:
$$C_{(a)} \underset{k_{-3}}{\overset{k_{+3}}{\rightleftharpoons}} Graphene_{(a)} \quad \Delta H_{form\_G} \cong -2.4 \text{ eV}^{[2a]}$$

Generally, the rate of adsorption:
$$r_{ad} = \frac{P}{\sqrt{2\pi mkT}} s_0 \exp\left(\frac{-E_a}{kT}\right) f(\theta_s)$$

Where P is the partial pressure of the reactant gas, m is the mass of the reactant molecule, $k$, Boltzmann's constant, $T$, temperature, $s_0$ is the initial sticking coefficient pre-exponential factor, and $f(\theta_s)$ is free surface coverage($\theta_s$), dependent sticking coefficient term, $E_a$, activation energy of the adsorption.

The rate of desorption is:
$$r_{des} = v_n \exp\left(\frac{-E_{des}}{kT}\right)[A]^n$$

Where n, is the order of the reaction, $v_n$ is n-th order vibrational frequency, $[A]$, the surface concentration of the adsorbate, and $E_{des}$, activation energy of desorption. The head of adsorption then can be calculated as: $\Delta H_{ad} = E_{des} - E_a$.

We note that the $CH_4$ decomposition reaction can be broken down into many intermediate steps, but it can be combined to give effective kinetic constants for the reaction. Several contributions[4] consider the intermediate steps of reaction (1) on transition metal surfaces in the usual growth conditions to be the following:

(1a) $CH_{4(g)} + S_{(Cu)} \underset{k_{-1a}}{\overset{k_{+1a}}{\rightleftharpoons}} CH_{4(a)}$

(1b) $CH_{4(a)} + S_{(Cu)} \underset{k_{-1b}}{\overset{k_{+1b}}{\rightleftharpoons}} CH_{3(a)} + H_{(a)}$

(1c) $CH_{3(a)} + S_{(Cu)} \underset{k_{-1c}}{\overset{k_{+1c}}{\rightleftharpoons}} CH_{2(a)} + H_{(a)}$

(1d) $CH_{2(a)} + S_{(Cu)} \underset{k_{-1d}}{\overset{k_{+1d}}{\rightleftharpoons}} CH_{(a)} + H_{(a)}$

(1e) $CH_{(a)} + S_{(Cu)} \underset{k_{-1e}}{\overset{k_{+1e}}{\rightleftharpoons}} C_{(a)} + H_{(a)}$

Here, we do not consider any side reactions that leads to direct formation of dimers and its derivatives (C$_2$, C$_2$H$_x$, etc.) from CH$_x$'s and we consider the above reactions to be the general reaction pathway for the dissociative adsorption of CH$_4$[4].

For the reaction (1a), the rates of adsorption ($r_{+1a}$) and desorption ($r_{-1a}$), and the equilibrium constant ($K_{1a}$) are:

$$r_{+1a} = k_{+1a} P_{CH_4} (\theta_S \rho_s)$$

$$r_{-1a} = k_{-1a} (\theta_{CH_4} \rho_s)$$

$$k_{+1a} P_{CH_4} (\theta_S \rho_s) = k_{-1a} [CH_4]_{eq}$$

$$\frac{k_{+1a}}{k_{-1a}} = \frac{[CH_4]_{eq}}{P_{CH_4} (\theta_S \rho_s)} = K_{1a}$$

$$K_{1a} = \frac{1}{\rho_s v_{-1a} \sqrt{2\pi m_{CH_4} kT}} \exp\left(\frac{-\Delta H_{1a}}{kT}\right)$$

And for the reaction (1b),

$$r_{+1b} = k_{+1b} (\theta_{CH_4} \rho_s)(\theta_S \rho_s)$$

$$r_{-1b} = k_{-1b} (\theta_{CH_3} \rho_s)(\theta_H \rho_s)$$

$$k_{+1b} [CH_4]_{eq} (\theta_S \rho_s) = k_{-1b} [CH_3]_{eq} [H]_{eq}$$

$$\frac{k_{+1b}}{k_{-1b}} = K_{1b} = \frac{[CH_3]_{eq} [H]_{eq}}{[CH_4]_{eq} (\theta_S \rho_s)}$$

$$K_{1b} = \frac{v_{+1b}}{v_{-1b}} \exp\left(\frac{-\Delta H_{1\_2}}{kT}\right)$$

Combining the reactions (1a and 1b), the effective equilibrium constant, $K_{1(a+b)}$ is:

$$K_{1(a+b)} = \frac{[CH_3]_{eq} [H]_{eq}}{P_{CH4} (\theta_S \rho_s)^2} = \frac{K_{1b} [CH_4]_{eq} (\theta_S \rho_s)}{\frac{[CH_4]_{eq}}{K_{1a}} (\theta_S \rho_s)} = K_{1a} K_{1b}$$

$$K_{1a}K_{1b} = \frac{v_{+1b}}{\rho_s v_{-1a} v_{-1b} \sqrt{2\pi m_{CH_4} kT}} \exp\left(\frac{-(\Delta H_{1b} + \Delta H_{1a})}{kT}\right)$$

$$= \frac{1}{\rho_s v_{1(a+b)} \sqrt{2\pi m_{CH_4} kT}} \exp\left(\frac{-\Delta H_{1(a+b)}}{kT}\right)$$

with the effective vibrational frequency factor, $v_{1(a+b)} = \frac{v_{-1a} v_{-1b}}{v_{+1b}}$,

and overall enthalpy, $\Delta H_{1(a+b)} = \Delta H_{1b} + \Delta H_{1a}$.

In this manner, the kinetic parameters of intermediate steps can therefore be reduced to a single set of effective constants for the overall CH$_4$ decomposition reaction (1) such that:

$$K_1 = K_{1(a+b+c+d+e)} = K_{1a}K_{1b}K_{1c}K_{1d}K_{1e}$$

$$= \frac{1}{\rho_s v_1 \sqrt{2\pi m_{CH_4} kT}} \exp\left(\frac{-\Delta H_{ad\_CH_4}}{kT}\right)$$

with $v_1 = v_{1(a+b+c+d+e)} = \frac{v_{-1a} v_{-1b} v_{-1c} v_{-1d} v_{-1e}}{v_{+1b} v_{+1c} v_{+1d} v_{+1b}}$

and $\Delta H_{ad\_CH_4} = \Delta H_{1(a+b+c+d+e)} = \Delta H_{1a} + \Delta H_{1b} + \Delta H_{1c} + \Delta H_{1d} + \Delta H_{1e}$.

Hence, balancing rates of adsorption and desorption for the overall methane decomposition reaction:

$$r_{+1} = k_{+1} P_{CH_4} (\theta_S \rho_s)^5$$

$$r_{-1} = k_{-1} [C][H]^4$$

$$k_{+1} P_{CH_4} (\theta_S \rho_s)^5 = k_{-1} [C]_{eq} [H]_{eq}^4$$

$$\frac{k_{+1}}{k_{-1}} = \frac{[C]_{eq} [H]_{eq}^4}{P_{CH_4} (\theta_S \rho_s)^5} = \frac{(\theta_C \rho_s)(\theta_H \rho_s)^4}{P_{CH_4} (\theta_S \rho_s)^5 (1-\theta_G)^5} = \frac{(\theta_C)(\theta_H)^4}{P_{CH_4} (\theta_S)^5 (1-\theta_G)^5} = K_1$$

$$K_1 = \frac{1}{\rho_s v_1 \sqrt{2\pi m_{CH_4} kT}} \exp\left(\frac{-\Delta H_{ad\_CH_4}}{kT}\right)$$

where we set $v_1 = 10^{13}$ s$^{-1}$, which is a generally used value of a vibrational frequency when the experimental value is not known[5], $m_{CH_4} = 2.67 \times 10^{-23}$ kg, and $\rho_{s\_Cu} \sim 1.53 \times 10^{19}$ m$^{-2}$.

Here, we use the assumption of the Langmuir adsorption mechanism for the sticking coefficient:

$$s_0 f(\theta_s) = \theta_s = 1 - \theta_H - \theta_C - \theta_G$$

Generally, $[A] = \theta_A \rho_s$, where $\theta_A$ is the overall coverage of the adsorbed species A, and $\rho_s$ is the concentration of surface sites. However, due to the presence of graphene the effective overall free area is decreased by the area coverage of graphene which effectively increases the surface concentration of the adsorbed species. Thus, we assume that C and H adatoms competitively bind only to the available surface that is not covered by graphene so that the surface concentrations of carbon and hydrogen can be expressed in terms of carbon, hydrogen, graphene coverage as:

$$[C] = \frac{\theta_C \rho_s}{1-\theta_G}, \text{ and } [H] = \frac{\theta_H \rho_s}{1-\theta_G}$$

Similarly, for the reaction (2), the rates of hydrogen desorption and adsorption are:

$$r_{+2} = k_{+2}[H]^4$$

$$r_{-2} = k_{-2}(P_{H_2})^2 (\theta_S \rho_s)^4$$

Balance the two rates to obtain the relationship at the equilibrium:

$$k_{+2}[H]_{eq}^4 = \frac{k_{+2}(\theta_H \rho_s)^4}{(1-\theta_G)^4} = k_{-2}(P_{H_2})^2 (\theta_S \rho_s)^4$$

$$\frac{k_{+2}}{k_{-2}} = \frac{(P_{H_2})^2 (\theta_S)^4 (1-\theta_G)^4}{(\theta_H)^4} = K_2$$

$$K_2 = \rho_s^2 v_2^2 (2\pi m_{H_2} kT) \exp\left(\frac{2\Delta H_{ad\_H_2}}{kT}\right)$$

where $v_2 = 10^{-13}$ s, and $m_{H_2} = 3.347 \times 10^{-24}$ kg.

For the graphene formation from adsorbed carbon, the rates of attachment ($r_{+3}$) and detachment ($r_{-3}$) of the carbon adatoms per unit length of graphene phase boundary are to be balanced for the steady state. According to the theory of 2D crystallization kinetics[6]:

$$r_{+3} = k_{+3}[C] = a_{cu} v_{+3\_Cu} \exp\left(\frac{-E_{att}}{kT}\right)[C]$$

$$r_{-3} = k_{-3} = \frac{v_{-3\_G}}{a_G} \exp\left(\frac{-E_{det}}{kT}\right)$$

$$k_{+3}[C]_{eq} = k_{-3}$$

$$\frac{k_{+3}}{k_{-3}} = \frac{1}{[C]_{eq}} = \frac{1-\theta_G}{\theta_C \rho_s} = K_3$$

$$K_3 = \frac{a_{cu} a_G v_{+3\_Cu}}{v_{-3\_G}} \exp\left(\frac{-\Delta H_{form\_G}}{kT}\right)$$

Here, graphene interatomic spacing, $a_G = 1.42 \times 10^{-10}$ m and $a_{Cu} = 2.3 \times 10^{-10}$ m. $v_{+3\_Cu}$ is related to the vibrational frequency of Cu, $v_{-3\_G}$ is related to the vibrational frequency of graphene. If $v_{+3\_Cu} \approx v_{-3\_G}$, the approximate value of the pre-exponential factor is in the order of $10^{-20}$ m$^2$.

Then,

$$\theta_C = \frac{K_1 K_2 P_{CH_4} \theta_S (1-\theta_G)}{(P_{H_2})^2} = \frac{(1-\theta_G)}{K_3 \rho_s}, \quad \theta_S = \frac{(P_{H_2})^2}{K_1 K_2 K_3 \rho_s P_{CH_4}}, \quad \text{and} \quad \theta_H = \frac{(P_{H_2})^{\frac{1}{2}} \theta_S (1-\theta_G)}{(K_2)^{\frac{1}{4}}}$$

Using the relationship, $\theta_H + \theta_C + \theta_S + \theta_G = 1$ and solving the system of equations:

$$\theta_G = 1 - \frac{K_2^{\frac{1}{4}} (P_{H_2})^2}{K_1 K_2^{\frac{5}{4}} K_3 P_{CH_4} \rho_s - K_1 K_2^{\frac{5}{4}} P_{CH_4} - (P_{H_2})^{\frac{5}{2}}}$$

In the typical experimental conditions (P$_{CH4}$, P$_{H2}$ < 1 MPa, T = 300 K – 1080 K), $\theta_S \gg \theta_C, \theta_H$.

Thus, $\frac{(P_{H_2})^2}{K_1 K_2 K_3 \rho_s P_{CH_4}} = \theta_S$ is the dominant term giving rise to exponential behavior with apparent activation energy of $\Delta H_{ad\_CH4} - 2\Delta H_{ad\_H2} + \Delta H_{form\_G}$.

Fixing the activation energies as given by the literature values and $v_1, v_2 = 10^{-13}$ s and using the pre-exponential factor of K$_3$ as the only fitting parameter, the curve fitting (figure 1c of the manuscript) was performed on our experimental data with $\frac{a_{cu} a_G v_{+3\_Cu}}{v_{-3\_G}} = 6.3 \times 10^{-18}$ m$^{-2}$

This is a reasonable value as the vibrational factors can vary over a few orders of magnitude.[5]

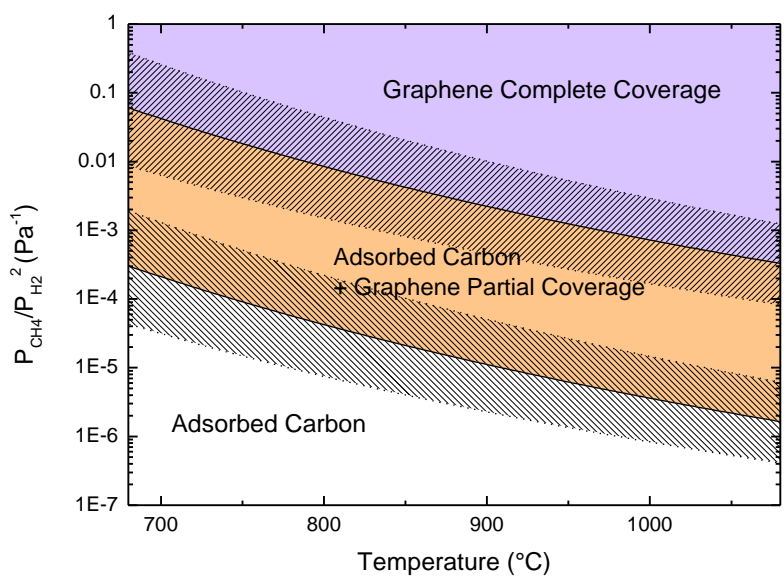

Figure S1. Contour diagram for complete graphene coverage ($\theta_G > 0.995$), adsorbed carbon ($\theta_G = 0$), and partial graphene coverage ($0 < \theta_G < 0.995$) with error bounds (shaded areas) considering 5 % error in the pre-exponential factors, 10% error in pressure values, and 14% error in the energy values (estimated based on the curve fitting of figure 1c).

**Supplemental Data References**